\documentclass[prl,twocolumn,superscriptaddress]{revtex4-1}
\usepackage{amssymb}
\usepackage{amsfonts}
\usepackage{bbm}
\usepackage{mathrsfs}
\usepackage{graphics,graphicx,epsfig,bm,amsmath,amsthm,amssymb}
\usepackage{bm}
\usepackage{bbm}
\usepackage{longtable}
\usepackage{multirow}
\usepackage{array}
\usepackage{color}
\usepackage{cases}
\usepackage{booktabs }
\usepackage[usenames,dvipsnames]{xcolor}

\usepackage{xcolor}

\usepackage{float}
\usepackage[a4paper,colorlinks=true,
linkcolor=blue,citecolor=blue,
pdfauthor={ },
pdftitle={ },
pdfsubject={ },
pdfkeywords={ }]{hyperref}

\begin{document}

\title{A programmable two-qubit solid-state quantum processor under ambient conditions}

\author{Yang Wu}
\affiliation{CAS Key Laboratory of Microscale Magnetic Resonance and Department of Modern Physics, University of Science and Technology of China, Hefei 230026, China}
\affiliation{Synergetic Innovation Center of Quantum Information and Quantum Physics, University of Science and Technology of China, Hefei 230026, China}
\affiliation{Hefei National Laboratory for Physical Sciences at the Microscale, University of Science and Technology of China, Hefei 230026, China}

\author{Ya Wang}
\affiliation{CAS Key Laboratory of Microscale Magnetic Resonance and Department of Modern Physics, University of Science and Technology of China, Hefei 230026, China}
\affiliation{Synergetic Innovation Center of Quantum Information and Quantum Physics, University of Science and Technology of China, Hefei 230026, China}
\affiliation{Hefei National Laboratory for Physical Sciences at the Microscale, University of Science and Technology of China, Hefei 230026, China}

\author{Xi Qin}
\affiliation{CAS Key Laboratory of Microscale Magnetic Resonance and Department of Modern Physics, University of Science and Technology of China, Hefei 230026, China}
\affiliation{Synergetic Innovation Center of Quantum Information and Quantum Physics, University of Science and Technology of China, Hefei 230026, China}
\affiliation{Hefei National Laboratory for Physical Sciences at the Microscale, University of Science and Technology of China, Hefei 230026, China}

\author{Xing Rong}
\email{xrong@ustc.edu.cn}
\affiliation{CAS Key Laboratory of Microscale Magnetic Resonance and Department of Modern Physics, University of Science and Technology of China, Hefei 230026, China}
\affiliation{Synergetic Innovation Center of Quantum Information and Quantum Physics, University of Science and Technology of China, Hefei 230026, China}
\affiliation{Hefei National Laboratory for Physical Sciences at the Microscale, University of Science and Technology of China, Hefei 230026, China}

\author{Jiangfeng Du}
\email{djf@ustc.edu.cn}
\affiliation{CAS Key Laboratory of Microscale Magnetic Resonance and Department of Modern Physics, University of Science and Technology of China, Hefei 230026, China}
\affiliation{Synergetic Innovation Center of Quantum Information and Quantum Physics, University of Science and Technology of China, Hefei 230026, China}
\affiliation{Hefei National Laboratory for Physical Sciences at the Microscale, University of Science and Technology of China, Hefei 230026, China}

\date{\today}

\begin{abstract}
Quantum computers, which take advantage of the superposition and entanglement of physical states, could outperform their classical counterparts in solving problems with technological impact, such as factoring large numbers and searching databases.
A quantum processor executes algorithms by applying a programmable sequence of gates to an initialized state of qubits, which coherently evolves into a final state containing the result of the computation.
Although quantum processors with a few qubits have been demonstrated on multiple quantum computing platforms, realization of solid-state programmable quantum processor under ambient conditions remains elusive.
Here we report a programable quantum processor that can be programmed with fifteen parameters to realize arbitrary unitary transformations on two spin-qubits in a nitrogen-vacancy (NV) center in diamond.
We implemented the Deutsch-Jozsa and Grover search algorithms with average success rates above 80\%.
This programmable two-qubit processor could form a core component of a large-scale quantum processor, and the methods used here are suitable for such a device.
\end{abstract}
%\pacs{Valid PACS appear here}

\maketitle

The versatility of computers come from changing the problem to be solved to reconfiguring inputs, that is, reprogramming it.
In a classical computer, a program is ultimately decomposed into sequences of
operations implemented with logic gates.
A similar decomposition exists for quantum processors\cite{PRSL_DJ_1989, PRA_Barenco, PRL_Bremner}.
Arbitrary operations on a multi-qubit system can be broken down into sequences of quantum gates.
Reconfiguring these gate sequences provides the flexibility to implement a variety of algorithms without altering the hardware.
As with its classical counterpart, a programmable quantum processor is more
versatile than one designed for a fixed task.
%Building such a quantum processors is challenging because of the simultaneously requirements that are in conflict: state preparation, long coherence times, universal gate operations and qubit readout.
Programmable quantum processors based on a few qubits have been demonstrated using trapped ions\cite{NatPhys_Wineland, Nature_Monroe}, superconducting qubits\cite{Nature_Schoelkopf} and quantum-dot-based qubits\cite{Nature_Vandersypen}, but the experimental realization of a solid-state programable quantum processor at room temperature remains elusive.
Here, we demonstrate that with the electron spin and $^{14}\text{N}$ nuclear spin of NV center in diamond one can combine initialization, readout, single- and two-qubit gates to form a programmable quantum processor that can perform quantum algorithms.

\begin{figure}\centering
\includegraphics[width=1.0\columnwidth]{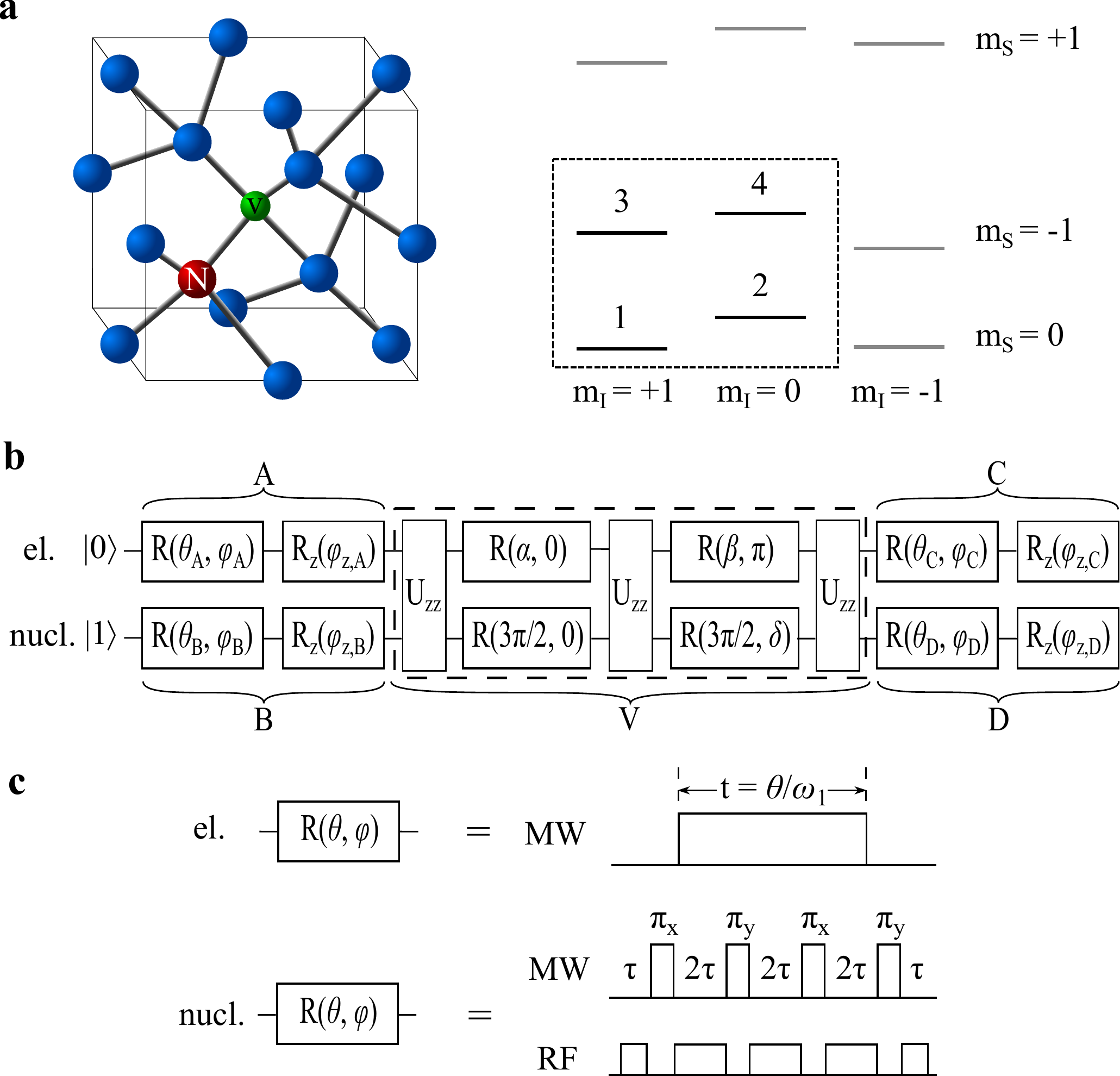}
\caption{(color online). Two-qubit programmable quantum processor in NV center. (a) Schematic atomic structure and energy levels of the NV center.
The experiments are implemented on the two-qubit system composed of four energy levels within the dashed box (i.e. $|m_S=0, m_I=+1\rangle$, $|m_S=-1, m_I=+1\rangle$, $|m_S=0, m_I=0\rangle$, and $|m_S=-1, m_I=0\rangle$ denoted by $|1\rangle$, $|2\rangle$, $|3\rangle$ and $|4\rangle$).
(b) Circuit diagrams for arbitrary unitary transformations.
The brackets highlight the decomposition of arbitrary two-qubit operation into $(C\otimes D)\cdot V \cdot (A\otimes B)$.
$R(\theta,\varphi)$ corresponds to a rotation angle $\theta$ around the axis in the XY plane.
$\varphi$ denotes the angle between the axis and the X axis. $R_z(\varphi_z)$ represents a rotation angle $\varphi_z$ around the Z axis.
The entangling two-qubit gate $U_{zz}$ is $\exp(i\pi S_z\otimes I_z)$.
(c) Realization of the single-qubit gates.
The single-qubit rotations on the electron spin are realized by hard microwave pulses with fixed strength $\omega_1 = 40~$MHz.
The duration of the pules corresponds to the rotation angle $\theta$.
The single-qubit rotations on the nuclear spin are realized by decoherence-protected gates consisting of dynamical decoupling of the electron spin and nuclear spin driving during the time interval.
The strength of the nuclear spin driving is determined by the rotation angle.
The time $\tau$ corresponds to the interaction between electron spin and nuclear spin.
}
    \label{Fig1}
\end{figure}

NV centers in diamond have emerged as one of the most promising candidates for implementing quantum technologies because they exhibit long coherence time and universal quantum gates with fault-tolerant control fidelity\cite{NatCommun_Du}.
As depicted in Fig.~\ref{Fig1}(a), the NV center consists of a substitutional nitrogen atom with an adjacent vacancy site in the diamond crystal lattice.
The ground state of NV center is an electron spin triplet state with three sublevels $|m_S=0\rangle$ and $|m_S=\pm1\rangle$.
The degeneracy between the $|m_S=+1\rangle$ and $|m_S=-1\rangle$ states is removed by applying a static magnetic field of about 500 G along the NV symmetry axis ([1 1 1] crystal axis).
Under such a magnetic field, the spin state of the NV center is effectively polarized to $|m_S=0, m_I=+1\rangle$ when a 532 nm laser pulse is applied\cite{PRL_Wrachtrup}.
The two-qubit quantum system is composed of $|m_S=0, m_I=+1\rangle$, $|m_S=0, m_I=0\rangle$, $|m_S=-1, m_I=+1\rangle$, and $|m_S=-1, m_I=0\rangle$ without considering the other spin levels.
The four energy levels are denoted by $|1\rangle$, $|2\rangle$, $|3\rangle$ and $|4\rangle$ as shown in Fig.~\ref{Fig1}(a).
%Microwave pulses driving electron spin transition $|m_S=0\rangle$ to $|m_S=-1\rangle$ and radio-frequency pulses driving nuclear spin transition $|m_I=+1\rangle$ to $|m_I=0\rangle$ are utilized to manipulate the spin states.
%The $|m_S=+1\rangle$ electron spin level and $|m_I=-1\rangle$ nuclear spin level remain idle due to large detuning.

Arbitrary single-qubit gates combined with applications of a maximally entangling two-qubit gate are sufficient for realization of universal two-qubit unitary operation.
Our choice of a universal gate library consists of single-qubit gates and an entangling two-qubit gate, $U_{zz}=\exp(i\pi S_z\otimes I_z)$, which can be realized by the free evolution under the hyperfine coupling between electron spin and nuclear spin.
$S_z$ and $I_z$ are the electron and nuclear spin operators, respectively.
We decompose a given unitary operation $U$ into $U=(C\otimes D)\cdot V \cdot (A\otimes B)$ as shown in Fig.~\ref{Fig1}b.
The decomposition follows a three-step procedure analogous to that in Ref.~\onlinecite{NatPhys_Wineland}.
First, we match matrix eigenvalues of $V$ and $U$ to find the three parameters $\alpha$, $\beta$ and $\delta$ shown in the dashed rectangle in Fig.~\ref{Fig1}(b).
Second, we calculate the four remaining single-qubit gates, $A$, $B$, $C$ and $D$, to fulfill the equation $U=(C\otimes D)\cdot V \cdot (A\otimes B)$.
Finally, we parameterize the single-qubit gates by the decomposition $R_z(\varphi_z)\cdot R(\theta,\varphi)$.
$R(\theta,\varphi)=\exp[-i\theta(\cos\varphi~S_x+\sin\varphi~S_y)]$, corresponds to a rotation of angle $\theta$ around the axis in the XY plane.
$\varphi$ denotes the angle between the axis of rotation and the X axis.
$R_z(\varphi_z)$ represents a rotation of angle $\varphi_z$ around the Z axis.
After the procedure discussed above, any two-qubit unitary transformation can be realized using the universal quantum circuit with fifteen parameters, which are $\alpha$, $\beta$, $\delta$, $\theta_A$, $\varphi_A$, $\varphi_{z,A}$, $\theta_B$, $\varphi_B$, $\varphi_{z,B}$, $\theta_C$, $\varphi_C$, $\varphi_{z,C}$, $\theta_D$, $\varphi_D$ and $\varphi_{z,D}$.

In our quantum processor the two-qubit entangling gate $U_{zz}$ is achieved by the free evolution under the hyperfine coupling between electron spin and nuclear spin.
The hyperfine coupling is characterized by Hamiltonian $H_{hf} = 2\pi A S_z I_z$ with the hyperfine coupling strength $A = -2.16 $ MHz.
So the time duration of $U_{zz}$ is 231.5 ns.
The single-qubit rotations on the electron spin are realized by microwave (MW) pulses with fixed strength of $\omega_{1MW}=40$ MHz which is large compared to the hyperfine coupling $A$.
This flips the electron spin practically independent of the nuclear spin state.
The length of the MW pulse $t$ is determined by the rotation angle $\theta$ as shown in the upper panel in Fig.~\ref{Fig1}(c).
Since the electron spin and nuclear spin evolve and decohere at very different rates, decoherence-protected gates are implemented to realize the single nuclear spin rotation\cite{Nature_Hanson} as shown in the lower panel in Fig.~\ref{Fig1}(c).
The XY-4 sequence is applied on the electron spin to protect the coherence.
The nuclear spin is driven during the time between the decoupling pulses.
A rotation of the nuclear spin, which is independent of the electron spin state, is constructed by choosing $\tau=2n\pi/A$ with integer $n$.
Here we fixed $\tau=2777.8$ ns with $n=6$ and the strength of the radio-frequency (RF) pulses is calculated by $\omega_{1RF}=\theta/4\tau$.
The rotation around the Z axis $R_z(\varphi_z)$ is realized by adding a phase to the drive field for all subsequent gates.
%The electron spin is dynamically protected when the nuclear spin is driven even though the evolution time exceeds the electron spin dephasing time.

Our experiment was implemented on an NV center in $[100]$ face bulk diamond.
The nitrogen concentration in the diamond was less than 5 ppb and the abundance of $^{13}$C was at the natural level of 1.1\%.
The diamond was mounted on a home-built confocal setup.
Spin-state initialization and detection of the NV center was realized with a 532 nm green laser controlled by an acousto-optic modulator (ISOMET, power leakage ratio $\sim$1/1000).
To preserve the NV center's longitudinal relaxation time from laser leakage effects, the laser beam was passed twice through the acousto-optic modulator before going through an oil objective (Olympus, PLAPON 60*O, NA 1.42).
The phonon sideband fluorescence (wavelength, 650-800nm) went through the same oil objective and was collected by an avalanche photodiode (Perkin Elmer, SPCM-AQRH-14) with a counter card.
A solid immersion lens (SIL) is created around the NV center to increase the fluorescence collection efficiency.
The magnetic field was provided by a permanent magnet and aligned by monitoring the variation of fluorescence counts.
The state of the two-qubit solid-state quantum processor can be effectively polarized to $|m_S=0, m_I=+1\rangle$ with 532 nm laser pumping when a static magnetic field of about 500 G is applied along the NV symmetry axis.
The polarization of electron spin and nuclear spin are 95\% and 98\%, respectively.
The spin states were manipulated with microwave and radio-frequency pulses.
The microwave and radio-frequency pulses were generated by an arbitrary waveform generator (Keysight M8190A), amplified individually with power amplifiers (Mini Circuits ZHL-30W-252-S+ for microwave pulses and LZY-22+ for radio-frequency pulses).
A broadband coplanar waveguide with $15~$GHz bandwidth was designed and fabricated to feed the microwave.
The radio-frequency pulses were carried by a home-built coil to suppress their thermal effects on the electron spin.

\begin{table*}[!h]\centering  % 表居中
%\textrm{\\}
\renewcommand{\multirowsetup}{\centering}
%\begin{tabular}{c|c|ccccccccccccccc}  % {lccc} 表示各列元素对齐方式，
\begin{tabular}{c|c|cp{0.75cm}<{\centering}p{0.75cm}<{\centering}cccccccccccc}  % {lccc} 表示各列元素对齐方式，
%\begin{tabular}{c|c|p{0.75cm}<{\centering}p{0.75cm}<{\centering}p{0.75cm}<{\centering}p{0.75cm}<{\centering}p{0.75cm}<{\centering}p{0.75cm}<{\centering}p{1cm}<{\centering}p{0.75cm}<{\centering}p{0.75cm}<{\centering}p{0.75cm}<{\centering}p{0.75cm}<{\centering}p{0.75cm}<{\centering}p{0.75cm}<{\centering}p{0.75cm}<{\centering}p{0.75cm}<{\centering}}  % {lccc} 表示各列元素对齐方式，left-l,right-r,center-c
\hline\hline
$U_i$ & $U_{meas}$ & $\alpha$ & $\beta$ & $\delta$ & $\theta_A$ & $\phi_A$ & $\theta_{z,A}$ & $\theta_B$ & $\phi_B$ & $\theta_{z,B}$ & $\theta_C$ & $\phi_C$ & $\theta_{z,C}$ & $\theta_D$ & $\phi_D$ & $\theta_{z,D}$ \\  \hline
\multirow{3}{*}{$U_1$} & $I$ & 0 & 0 & 0 & 0.785 & 0 & 1.571 &         1.571 & -1.571 & 3.927 & 0.785 & -1.571 & 4.712 & 0.785 & -1.571 & 1.571   \\  %\cline{2-5} % \hline 在此行下面画一横线
& $\pi_e$ & 0 & 0 & 0 & 0.785 & 0 & 1.571 & 1.571 & -1.571 & 3.927 & 3.927 & -1.571 & 4.712 & 0.785 & -1.571 & 1.571   \\
& $\pi_n$ & 0 & 0 & 0 & 0.785 & 0 & 1.571 & 1.571 & -1.571 & 3.927 & 0.785 & -1.571 & 4.712 & 2.356 & -1.571 & 1.571   \\  \hline
\multirow{3}{*}{$U_2$} & $I$ & 1.571 & 0 & 0 & 4.712 & -0.785 & 0.785 & 4.712 & -1.571 & 1.571 & 1.571 & 0 & 0.785 & 1.571 & 1.571 & 4.712  \\
& $\pi_e$ & 1.571 & 0 & 0 & 4.712 & -0.785 & 0.785 & 4.712 & -1.571 & 1.571 & 1.571 & -3.142 & 5.498 & 1.571 & 1.571 & 4.712  \\
& $\pi_n$ & 1.571 & 0 & 0 & 4.712 & -0.785 & 0.785 & 4.712 & -1.571 & 1.571 & 1.571 & 0 & 0.785 & 1.571 & -1.571 & 4.712  \\

\hline \hline

\end{tabular}
\caption{The parameters of the two-qubit programmable processor for Deutsch-Josza algorithm. The constant function and the balanced function are mapped onto the unitary operators $U_1 = I$ and $U_2 = \text{CNOT}$, respectively. The operation $U_{meas}$ indicate the measurement pulse after the algorithm, where $I$ denotes the identity and $\pi_e$ ($\pi_n$) denotes single-qubit $\pi$ pulse on electron (nuclear) spin.}
\label{DJ_para}
\end{table*}

\begin{figure}\centering
\includegraphics[width=1.0\columnwidth]{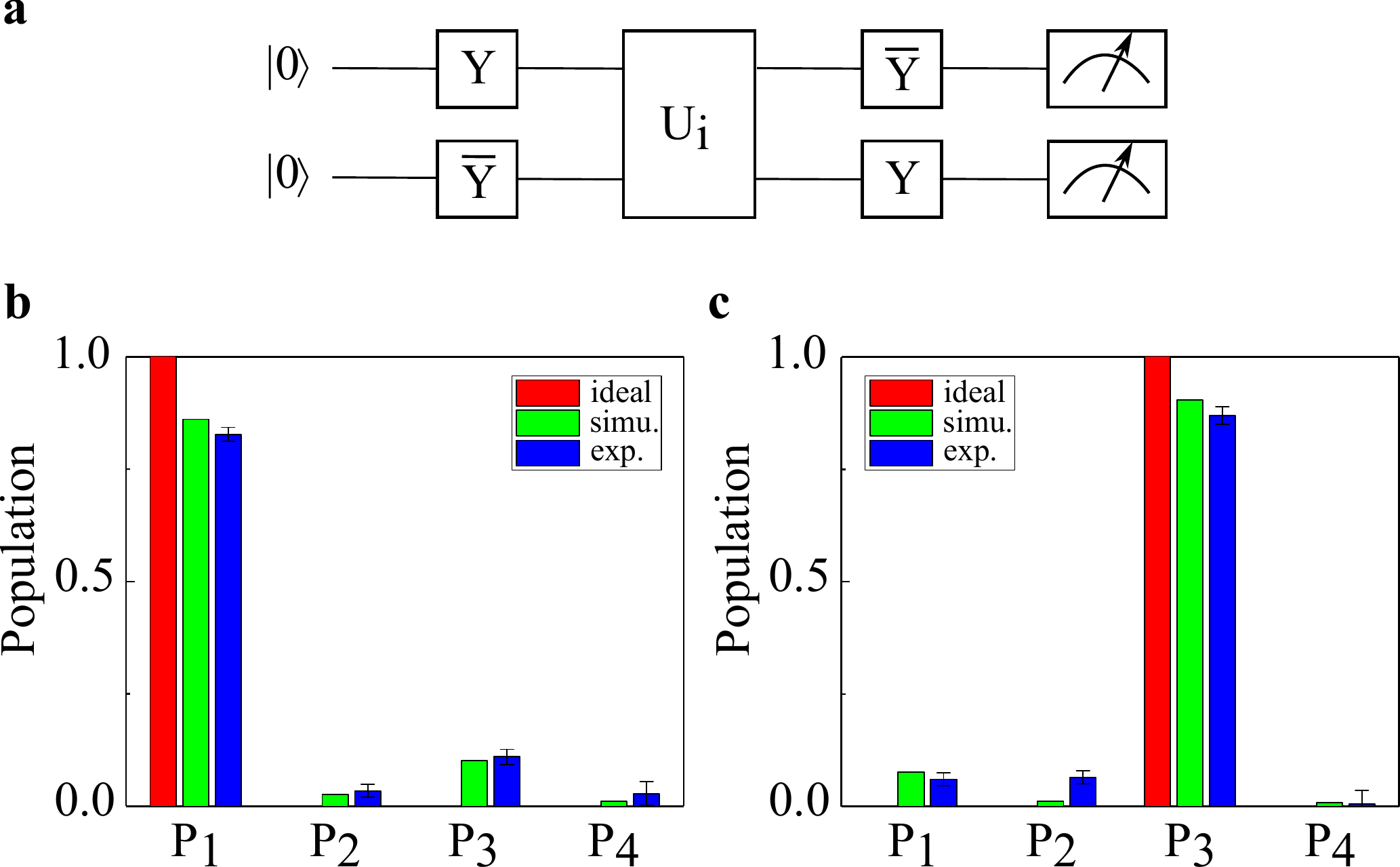}
\caption{(color online). Experimental demonstration of Deutsch-Josza algorithm. (a) Quantum circuit of Deutsch-Josza algorithm. $\text{Y}$($\overline{\text{Y}}$) indicate $\pi/2$($-\pi/2$) rotation around the $\text{Y}$ axis. The constant function and the balanced function are mapped onto the unitary operators $U_1 = I$ and $U_2 = \text{CNOT}$, respectively. (b) and (c) The experimental results of Deutsch-Josza algorithm with constant and balanced functions, respectively. $P_1$, $P_2$, $P_3$ and $P_4$ denoted the population of $|1\rangle$, $|2\rangle$, $|3\rangle$ and $|4\rangle$, respectively. The red bars are the ideal case. The green bars are the simulated case considering the imperfect polarization and the decoherence. The blue bars are the experiment results.
}
    \label{Fig2}
\end{figure}

To demonstrate the ability of the processor to generate arbitrary unitary transformations, we first implement the Deutsch-Jozsa algorithm\cite{PRSL_DJ} as an example.
It has been employed in different systems to demonstrate the exponential speedup in distinguishing constant from balanced functions with respect to the corresponding classical algorithm\cite{Nature_Chuang, PRL_DJ_Fazhan Shi, PRA_DJ_Chenyong Ju, PRB_DJ_Piermarocchi, PRL_DJ_Benson, PRL_DJ_Zeilinger, Nature_DJ_Blatt}.
A function that has an $n$-bit input and a 1-bit output ($f: {0, 1, 2,...,
2^n-1}\rightarrow{0, 1}$) is balanced when exactly half of the inputs result in the output 0 and the other half in the output 1, while a constant function
assumes a single value irrespective of the input.
In our setup, the constant function ($f(0)=f(1)=0$) and the balanced function ($f(0) = 0, f(1) = 1$) are mapped onto the unitary operators $U_i$ with $U_1 = I$ and $U_2 = \text{CNOT}$ as shown in the quantum circuit in Fig.~\ref{Fig2}(a), where the $I$ denotes the identity and the CNOT is the controlled-NOT gate.
A measurement of the populations, $P=\{P_1P_2P_3P_4\}$, of the final state determines if the function is constant or balanced.
To determine the level occupation probability of the final state, we measure the photoluminescence of the state.
And we also apply $\pi_e$ and $\pi_n$ pulses on the electron and nuclear spin respectively on the final state to flip the populations within the two-qubit system.
The population of each level can be calculated from the resulting photoluminescence of these states.
The entire quantum circuit combined with the flipping pulses can be treated as a unitary operator.
We decompose this unitary operator into the universal quantum circuit as shown in Fig.~\ref{Fig1}b following the three steps discussed above and obtain the 15 parameters as shown in Table~\ref{DJ_para}.
Different unitary operators are implemented using the universal quantum circuit with different sets of parameters.
%The unitary operator of the entire quantum circuit combined with the measurement pulses are programmed into the 15 parameters as shown in Table~\ref{DJ_para}.
%Changing unitary operator results in changes just on the parameters without any other alteration on the fixed quantum circuit and the structure of the qubits themselves.
Measurement of output $P=\{1000\}$ indicates a constant function, while $P=\{0010\}$ indicates a balanced function.
In Fig.~\ref{Fig2}(b, c) we show the measured population of each spin level of the final state after the Deutsch-Josza algorithm.
%The red bars indicate the ideal result, while the simulation case gives the result under the consideration of the decoherence of the electron spin and the imperfect polarization.
The red bars indicate the ideal result, while the green bar is the simulation result considering the decoherence of the electron spin and the imperfect polarization.
The experimental results (blue bars) are in good agreement with the simulated results.
The average success probability is 0.88(2) for constant and 0.93(2) for balanced functions.

\begin{figure}\centering
\includegraphics[width=1.0\columnwidth]{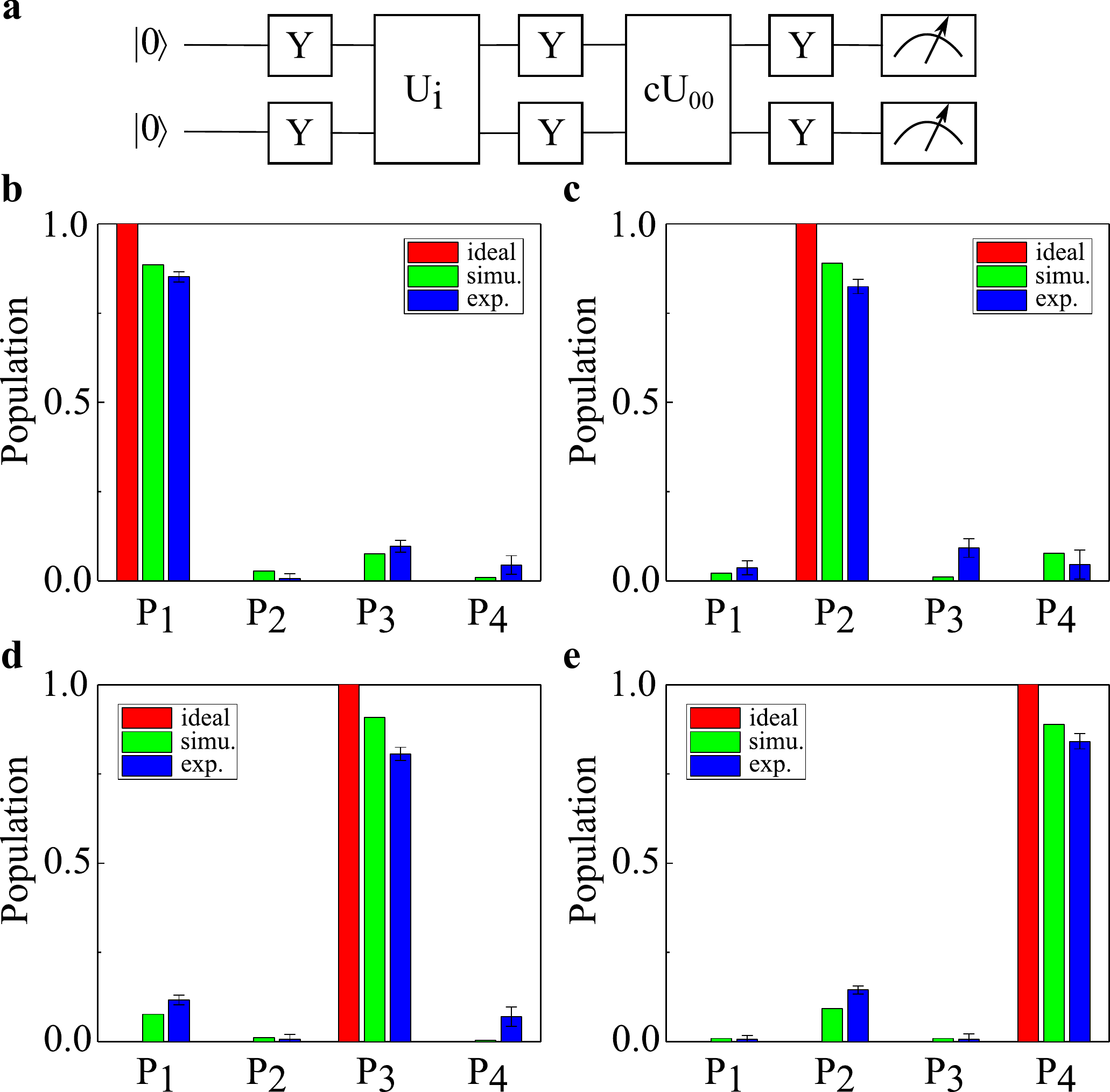}
\caption{(color online). Experimental demonstration of Grover search algorithm. (a) Quantum circuit of Grover search algorithm. Four possible function $f_{ij}(x)$ are mapped onto the oracle $U_i$ which is the C-Phase gate $cU_{ij}$. (b-e) The results of Grover search algorithm characterized by the population of each spin level of the final state. $P_1$, $P_2$, $P_3$ and $P_4$ denoted the population of $|1\rangle$, $|2\rangle$, $|3\rangle$ and $|4\rangle$, respectively. The red bars are the ideal case. The green bars are the simulated case considering the imperfect polarization and the decoherence. The blue bars are the experiment results.
}
    \label{Fig3}
\end{figure}

Now we show that another quantum algorithm, named Grover search algorithm, can be executed by our programable quantum processor by just adjusting the fifteen parameters. The  Grover’s search algorithm\cite{PRL_Grover} provides an optimal method for finding the unique input value $x_0$ of a function $f(x)$ that gives $f(x_0) = 1$ and $f(x) = 0$ for all other values of $x$.
It has been demonstrated in various systems\cite{Nature_Jones, PRL_Grover_Chuang, Nature_Gorver_Zeilinger, PRA_Grover_Monroe}.
In the two-qubit version of this algorithm there are four input values, $x \in (00, 01, 10, 11)$, resulting in four possible functions $f_{ij}(x)$,
with $i, j \in (0, 1)$.
These functions are mapped onto the C-Phase gate $cU_{ij}$ that encodes $f_{ij}(x)$ in a quantum phase, $cU_{ij}|x\rangle = (-1)^{f_{ij}(x)} |x\rangle$.
The C-Phase gate $cU_{ij}$ is denoted by the oracle $U_{i}$ as shown in the quantum circuit in Fig.~\ref{Fig3}(a).
The unitary operator of the four possible quantum circuit is programmed into the four sets of parameters as shown in Table~\ref{Grover_para}.
A measurement of the populations, $P=\{P_1P_2P_3P_4\}$, of the final state finds the state that has been marked.
In Fig.~\ref{Fig3}(b-e) we show the measured the population of each spin level of the final state after the Grover search algorithm of four possible oracle functions.
The experimental results (blue bars) are in good agreement with the simulated results (green bars).
The average success probability is 0.85(1), 0.82(2), 0.81(2) and 0.84(2) for four possible functions, respectively.

\begin{table*}[!h]\centering  % 表居中
%\textrm{\\}
\renewcommand{\multirowsetup}{\centering}
%\begin{tabular}{c|c|ccccccccccccccc}  % {lccc} 表示各列元素对齐方式，
\begin{tabular}{c|c|p{0.75cm}<{\centering}p{0.75cm}<{\centering}p{0.75cm}<{\centering}cccccccccccc}  % {lccc} 表示各列元素对齐方式，
%\begin{tabular}{c|c|p{0.75cm}<{\centering}p{0.75cm}<{\centering}p{0.75cm}<{\centering}p{0.75cm}<{\centering}p{0.75cm}<{\centering}p{0.75cm}<{\centering}p{1cm}<{\centering}p{0.75cm}<{\centering}p{0.75cm}<{\centering}p{0.75cm}<{\centering}p{0.75cm}<{\centering}p{0.75cm}<{\centering}p{0.75cm}<{\centering}p{0.75cm}<{\centering}p{0.75cm}<{\centering}}  % {lccc} 表示各列元素对齐方式，left-l,right-r,center-c
\hline\hline
$U_i$ & $U_{meas}$ & $\alpha$ & $\beta$ & $\delta$ & $\theta_A$ & $\phi_A$ & $\theta_{z,A}$ & $\theta_B$ & $\phi_B$ & $\theta_{z,B}$ & $\theta_C$ & $\phi_C$ & $\theta_{z,C}$ & $\theta_D$ & $\phi_D$ & $\theta_{z,D}$ \\  \hline
\multirow{3}{*}{$U_1$} & $I$ &0 & 0 & 0 & 0.785 & 1.571 & 0 & 1.571 & 0 & 2.356 & 0.785 & -1.571 & 4.712 & 0.785 & -1.571 & 1.571   \\
& $\pi_e$ &0 & 0 & 0 & 0.785 & 1.571 & 0 & 1.571 & 0 & 2.356 & 3.927 & -1.571 & 4.712 & 0.785 & -1.571 & 1.571   \\
& $\pi_n$ &0 & 0 & 0 & 0.785 & 1.571 & 0 & 1.571 & 0 & 2.356 & 0.785 & -1.571 & 4.712 & 2.356 & 1.571 & 1.571   \\ \hline

\multirow{3}{*}{$U_2$} & $I$ &0 & 0 & 0 & 0.785 & 1.571 & 0 & 1.571 & 3.142 & 2.356 & 0.785 & -1.571 & 4.712 & 0.785 & -1.571 & 1.571   \\
& $\pi_e$ &0 & 0 & 0 & 0.785 & 1.571 & 0 & 1.571 & 3.142 & 2.356 & 3.927 & -1.571 & 4.712 & 0.785 & -1.571 & 1.571   \\
& $\pi_n$ &0 & 0 & 0 & 0.785 & 1.571 & 0 & 1.571 & 3.142 & 2.356 & 0.785 & -1.571 & 4.712 & 2.356 & 1.571 & 1.571   \\ \hline

\multirow{3}{*}{$U_3$} & $I$ &0 & 0 & 0 & 3.927 & -1.571 & 3.142 & 1.571 & 0 & 2.356 & 0.785 & -1.571 & 4.712 & 0.785 & -1.571 & 1.571  \\
& $\pi_e$ &0 & 0 & 0 & 3.927 & -1.571 & 3.142 & 1.571 & 0 & 2.356 & 3.927 & -1.571 & 4.712 & 0.785 & -1.571 & 1.571   \\
& $\pi_n$ &0 & 0 & 0 & 3.927 & -1.571 & 3.142 & 1.571 & 0 & 2.356 & 0.785 & -1.571 & 4.712 & 2.356 & 1.571 & 1.571   \\ \hline

\multirow{3}{*}{$U_4$} & $I$ &0 & 0 & 0 & 3.927 & -1.571 & 3.142 & 1.571 & 3.142 & 2.356 & 0.785 & -1.571 & 4.712 & 0.785 & -1.571 & 1.571  \\
& $\pi_e$ &0 & 0 & 0 & 3.927 & -1.571 & 3.142 & 1.571 & 3.142 & 2.356 & 3.927 & -1.571 & 4.712 & 0.785 & -1.571 & 1.571   \\
& $\pi_n$ &0 & 0 & 0 & 3.927 & -1.571 & 3.142 & 1.571 & 3.142 & 2.356 & 0.785 & -1.571 & 4.712 & 2.356 & 1.571 & 1.571   \\
\hline \hline

\end{tabular}
\caption{The parameters of the two-qubit programmable processor for Grover search algorithm. Four possible function $f_{ij}(x)$ are mapped onto the oracle $U_i$. The operation $U_{meas}$ indicate the measurement pulse after the algorithm, where $I$ denotes the identity and $\pi_e$ ($\pi_n$) denotes single-qubit $\pi$ pulse on electron (nuclear) spin.}
\label{Grover_para}
\end{table*}

The algorithms presented here illustrate the computational flexibility provided by the solid-spin based quantum architecture at room temperature.
The programmable quantum processor is capable of implementing arbitrary unitary operation on two qubits using the universal quantum circuit with altering the parameters.
Substantial improvements could be made in the performance of the processor by using isotopically purified $^{12}\text{C}$, which would increase the coherence times of the qubits\cite{NatMat_Wrachtrup}.
Furthermore, dipolar coupling of electronic spins could mediate interactions between nuclear spins associated with different NV centers\cite{PRL_Retzker}, offering a potentially scalable platform for information processing\cite{NatComm_Lukin}.
Optical channels provide an alternate platform that is well suited to mediating interactions over macroscopic distances or in highly connected networks\cite{Nature_Monroe_07, Nature_Lukin}.
With these improvements quantum computers with multiple qubits and fidelities above the fault-tolerance threshold should be realizable.

We are grateful to D. Suter for valuable discussion. This work was supported by the National Key R$\&$D Program of China (Grants No. 2018YFA0306600, No. 2016YFB0501603 and 2017YFA0305000), the CAS (Grants No. GJJSTD20170001, No.QYZDY-SSW-SLH004 and No.QYZDB-SSW-SLH005) and Anhui Initiative in Quantum Information Technologies (Grant No. AHY050000). X.R. thanks the Youth Innovation Promotion Association of Chinese Academy of Sciences for their support. Y.W. thanks the projects supported by the NNSFC (Grants No. 11775209), the Innovative Program of Development Foundation of Hefei Center for Physical Science and Technology (Grants No. 2017FXCX005) and Project of Thousand Youth Talents.

Y. Wu and Y. Wang contributed equally to this work.

%\begin{addendum}

%\end{addendum}

%\newpage

\end{document}